\documentclass[aps,prb,twocolumn,superscriptaddress,showpacs,floatfix]{revtex4}

\usepackage{graphicx}
\usepackage{epstopdf}
\usepackage{dcolumn}
\usepackage{bm}

\begin{document}

\title {Structure and Magnetic Properties of the Pyrochlore Iridate Y$_2$Ir$_2$O$_7$}

\author{M. C. Shapiro}
\affiliation{Geballe Laboratory for Advanced Materials and Department of Applied Physics, Stanford University, Stanford, CA 94305, USA}
\affiliation{Stanford Institute of Energy and Materials Science, SLAC National Accelerator Laboratory, 2575 Sand Hill Road, Menlo Park, CA 94025, USA}

\author{S. C. Riggs}
\affiliation{Geballe Laboratory for Advanced Materials and Department of Applied Physics, Stanford University, Stanford, CA 94305, USA}
\affiliation{Stanford Institute of Energy and Materials Science, SLAC National Accelerator Laboratory, 2575 Sand Hill Road, Menlo Park, CA 94025, USA}

\author{M. B. Stone}
\affiliation{Neutron Sciences Directorate, Oak Ridge National
Laboratory, Oak Ridge, TN 37831, USA}

\author{C. R. de la Cruz}
\affiliation{Neutron Sciences Directorate, Oak Ridge National
Laboratory, Oak Ridge, TN 37831, USA}

\author{S. Chi}
\affiliation{Neutron Sciences Directorate, Oak Ridge National
Laboratory, Oak Ridge, TN 37831, USA}

\author{A. A. Podlesnyak}
\affiliation{Neutron Sciences Directorate, Oak Ridge National
Laboratory, Oak Ridge, TN 37831, USA}

\author{I. R. Fisher}
\affiliation{Geballe Laboratory for Advanced Materials and Department of Applied Physics, Stanford University, Stanford, CA 94305, USA}
\affiliation{Stanford Institute of Energy and Materials Science, SLAC National Accelerator Laboratory, 2575 Sand Hill Road, Menlo Park, CA 94025, USA}

\begin{abstract}
Neutron powder diffraction and inelastic measurements were performed examining the $5d$ pyrochlore Y$_2$Ir$_2$O$_7$. Temperature dependent measurements were performed between 3.4 K and 290 K, spanning the magnetic transition at 155 K.  No sign of any structural or disorder induced phase transition was observed over the entire temperature range.  In addition, no sign of magnetic long-range order was observed to within the sensitivity of the instrumentation. These measurements do not rule out long range magnetic order, but the neutron powder diffraction structural refinements do put an upper bound for the ordered iridium moment of $\sim$ 0.2 $\mu_B /$Ir (for a magnetic structure with wave vector $Q \neq 0$) or $\sim$ 0.5 $\mu_B /$Ir (for $Q = 0$).

\end{abstract}


\maketitle

\section{INTRODUCTION}

The rare earth iridate pyrochlores R$_2$Ir$_2$O$_7$ (R = Y and Pr-Lu) provide a fascinating opportunity to investigate the effects of spin-orbit coupling and geometric frustration on the magnetic and electronic properties of a correlated material. For R = Y and Nd-Lu, the materials exhibit a coupled magnetic and electronic transition \cite{taira01, yana01, hinatsu07} at a temperature $T_{mag}$ that depends on the rare earth ion. The effect must derive from the Ir electrons since it is observed not only for magnetic rare earths, but also for the non-magnetic elements Y and Lu. The magnetic properties are strongly hysteretic, and hence the transition was initially described in terms of freezing into a spin glass-like state \cite{taira01, yana01, hinatsu07}, but recent $\mu$SR measurements for Eu$_2$Ir$_2$O$_7$ indicate a homogeneous internal field and therefore long range magnetic order below $T_{mag}$. \cite{zhau11} The nature of the magnetically ordered state is currently unknown. 

The naturally occuring isotopes of Ir have a relatively large cross section for neutron absorption, but neutron diffraction measurements are nevertheless possible. Recent inelastic measurements for Nd$_2$Ir$_2$O$_7$ \cite{tomiyasu11} reveal splitting of the Nd crystal electric field (CEF) doublet ground state at $T_{mag}$ $\approx$ 33 K, implying the presence of a finite internal field due to order on the Ir sites. The Nd moments are found to order with a $Q = 0$ structure at a lower temperature $T_{\mathrm{Nd}}$ $\approx$ 15 K. The ordered moment is $M_{\mathrm{Nd}}$ = 2.4 $\pm$ 0.4 $\mu_B$, consistent with the proposed CEF scheme. Based on the small magnetic susceptibility, an all-in all-out structure was proposed for the Nd sublattice, in which the uniaxial anisotropy along the [111] directions is due to the doublet CEF ground state. \cite{tomiyasu11} However, no signature of long range order associated with the Ir moments was observed at $T_{mag}$. Recent $\mu$SR measurements for Nd$_2$Ir$_2$O$_7$ indicate the presence of a disordered magnetic state over an extended temperature range but confirm the presence of an ordered magnetic state at low temperatures. \cite{disseler11} The situation is further complicated due to uncertainty in the crystal structure, which is a crucial component of understanding the magnetic properties and the magnetic phase transition in any material. Raman scattering measurements for Nd$_2$Ir$_2$O$_7$ do not show any evidence for development of additional frequencies for $T < T_{mag}$, but both Sm$_2$Ir$_2$O$_7$ and Eu$_2$Ir$_2$O$_7$ show clear signatures of a reduction in the crystal symmetry at $T_{mag}$. \cite{hasegawa10} It is unclear what role this plays in reducing the geometric frustration inherent in the pyrochlore lattice. 

The nature of the magnetically ordered state of R$_2$Ir$_2$O$_7$ takes on an additional significance in light of recent electronic structure calculations. Tight binding calculations for the nonmagnetic structure indicate the possibility of a topological insulator for a wide range of values of the spin-orbit coupling and the Coulomb interaction \cite{pesin10}, although direct orbital overlap might affect this. \cite{krempa11} Significantly, long range magnetic order breaks time reversal symmetry and can profoundly change the nature of the electronic structure. Minimizing the energy for different possible magnetic structures, Wan and coworkers proposed an all-in all-out $Q = 0$ magnetic structure as a candidate ground state for the Ir sublattice, and showed that such a magnetic structure can result in exotic electronic phases, including a Weyl semimetal. \cite{ashvin11} Since these electronic states depend sensitively on both the crystal symmetry and also the magnetic structure, it is especially important to obtain measurements of the crystal and magnetic structure below $T_{mag}$. 

In this paper we present results of magnetization measurments and both elastic and inelastic powder neutron diffraction experiments for Y$_2$Ir$_2$O$_7$, chosen specifically since Y is a nonmagnetic ion. The measurements reveal no evidence for any change in the crystal structure through $T_{mag}$, and put an upper bound for the ordered iridium moment of $\sim$ 0.2 $\mu_B /$Ir (for a magnetic structure with wave vector $Q \neq 0$) or $\sim$ 0.5 $\mu_B /$Ir (for $Q = 0$).

\section{EXPERIMENTAL METHODS}
Mixtures of Y$_2$O$_3$ and IrO$_2$ with purities of 99.99$\%$ were ground in stoichiometric molar ratios, pelletized, and then heated in air at 1000 $^{\circ}$C for 100 hours. The resulting material was reground, pressed into pellets, and resintered at the same temperature for an additional 150 hours, with two intermediate regrindings. Powder X-ray diffraction measurements confirmed the phase purity of the resulting Y$_2$Ir$_2$O$_7$ to within the resolution of the measurement. 

Magnetization measurements were performed using a commercial Quantum Design Magnetic Properties Measurement System (MPMS) magnetometer.  Measurements were performed as a function of magnetic field and temperature following initial zero field-cooling (ZFC) and field-cooling (FC) thermal cycles. 

Neutron scattering measurements were performed using the HB2A powder diffractometer at the High Flux Isotope Reactor (HFIR) at the Oak Ridge National Laboratory.  The powder sample was mounted as a 0.5 mm thick powder in an annular aluminum sample can with a helium exchange gas.  The sample was mounted within a top-loading displex sample environment.  Neutrons with a wavelength of $\lambda = 1.5374$ {\AA} were scattered from the sample using a collimation of 12' monochromator - 31' sample - 6' detector.  Diffraction patterns were measured between $T = 3.4$ K and $T = 290$ K.  High-flux coarse-resolution measurements were also performed using the WAND diffractometer at HFIR.  The identical sample was used as the HB2A measurements with a $^4$He flow cryostat.  Neutrons with a wavelength of $\lambda = 1.4827$ {\AA} were measured for a minimum of 6 hours per sample temperature.

Inelastic neutron scattering measurements were performed using the CNCS and ARCS time-of-flight chopper spectrometers at the Spallation Neutron Source at the Oak Ridge National Laboratory. \cite{ehlers11, abernathy11}  The identical sample was used in these measurements as in the diffraction measurements.  Data were collected as a function of temperature and incident energy, $E_i$.

\section{RESULTS AND DISCUSSION}

\begin{figure}[tb]
\centerline{\includegraphics[width=3.75in]{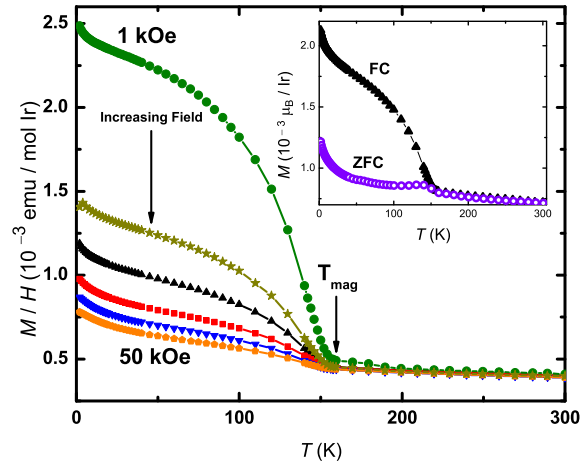}}
\caption{\label{fig:y2ir2o7mt}
(color online) Temperature dependence of the FC magnetization of Y$_2$Ir$_2$O$_7$ for applied magnetic fields of 1 kOe (green circles), 5 kOe (yellow stars), 10 kOe (black triangles), 20 kOe (red squares), 30 kOe (downward pointing blue triangles), and 50 kOe (orange pentagons). Magnetic fields refer both to the field applied during cooling and during measurement. Data have been normalized per mole of Ir (i.e., half of a formula unit). The magnetic transition at $T_{mag}$ $\approx$ 155 K is indicated. The inset compares the FC magnetization (black triangles) and ZFC magnetization (open purple circles) in units of $\mu_B /$Ir for a field of 10 kOe, revealing the substantial thermal hysteresis.}
\end{figure}

\begin{figure}[tb]
\centerline{\includegraphics[width=3.75in]{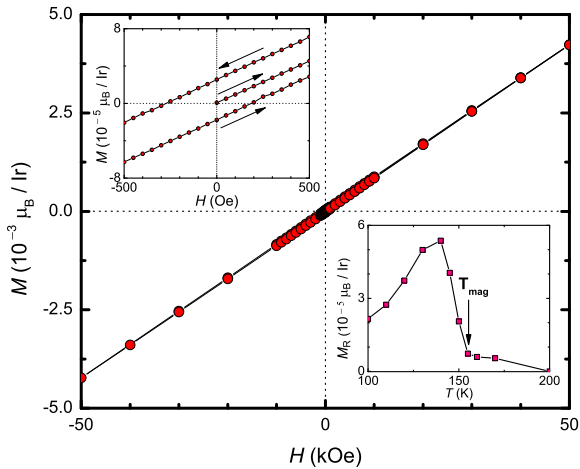}}
\caption{\label{fig:y2ir2o7mh}
(color online) Magnetization as a function of applied magnetic field at 100 K, well below $T_{mag}$.  The magnetization curve does not saturate up to $H = \pm$ 50 kOe.  Top left inset: Expanded view of the low field portion of the data shown in the main panel, revealing a small hysteresis.  Bottom right inset: Temperature dependence of the remanent magnetization $M_R$.  This value was calculated from the average of the absolute value of the zero-field intercepts from each magnetization loop.  $M_R$ steadily rises to a maximum at $T \approx$ 140 K, and then rapidly decreases as it passes through $T_{mag}$.  At slightly higher temperatures, $M_R$ goes to zero.}
\end{figure}

\begin{figure}[tb]
\centerline{\includegraphics[width=3.5in]{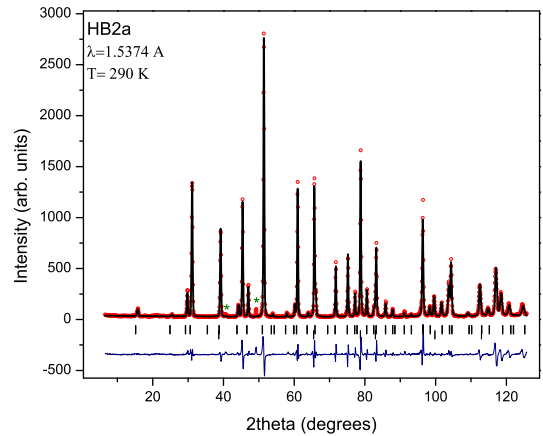}}
\caption{\label{fig:Ivstwotheta}
(color online) Powder diffraction data (red open circles) measured at $T = 290$ K using the HB2A diffractometer.  Solid heavy black line is the refined diffraction pattern as described in the text.  Vertical ticks are the indexed peak positions of the refinement, and the blue thin line is the difference in the measured and calculated diffraction pattern.  Green asterisks indicate the positions of a small temperature independent impurity phase.}
\end{figure}

Figure \ref{fig:y2ir2o7mt} shows the FC magnetization of Y$_2$Ir$_2$O$_7$ as a function of temperature for several different applied fields. The data clearly reveal a magnetic transition at $T_{mag}$ $\approx$ 155 K, consistent with previous results. \cite{maeno02, sato03, taira01} For temperatures above $T_{mag}$, the susceptibility appears to follow a linear temperature dependence, irrespective of the applied field, deviating from simple Curie-Weiss behavior. The inset of Figure \ref{fig:y2ir2o7mt} contrasts FC and ZFC magnetization data for $H =$ 10 kOe, revealing a large hysteretic difference below $T_{mag}$.  To further investigate this effect, magnetization measurements as a function of applied field were also made for temperatures spanning $T_{mag}$ (Figure \ref{fig:y2ir2o7mh}).  The magnetization shows no sign of saturating up to 50 kOe for all temperatures measured, but the data do reveal a clear hysteresis for temperatures below $T_{mag}$. The remanent magnetization $M_R$ is small (insets of Figure \ref{fig:y2ir2o7mh}) and rapidly drops at $T_{mag}$. Since the largest change in the magnetic hysteresis appears tied to the magnetic phase transition, it is unlikely to come from the small impurity phase seen in the powder neutron diffraction data (Figure \ref{fig:Ivstwotheta}). The origin of the hysteresis is not clear, but given the observation of an ordered magnetic state in the analogous compound Eu$_2$Ir$_2$O$_7$ via $\mu$SR measurements, it is presumably related either to changes in the domain population or possibly freezing of free spins at domain boundaries or other interfaces. In both cases, the reduction in $M_R$ at low temperatures would be related to thermal activation. The small magnitude of the remanent magnetization appears to rule out magnetic structures with a large net moment, such as two-in two-out and three-in one-out type configurations. A small hysteresis is still evident for temperatures just above $T_{mag}$, dropping to zero by 200 K, but the origin of this effect is unclear.

\begin{table*}[tp]
\centering
\begin{tabular}{| c | c | c|c  | c  | c  | l        | l        |l        | l        | l        | l       |}
  \hline
                Atom   & Ion  & Site &$x/a$  & $y/a$  & $z/a$  & $U_{11}$        & $U_{22}$        & $U_{33}$        & $U_{12}$        & $U_{13}$        & $U_{23}$    \\
  \hline
                  Y  &  Y$^{3+}$  & 16d & 0.5        & 0.5   & 0.5    & 0.00451(6) & 0.00451(6) & 0.00451(6) & -0.00097(7) & -0.00097(7) & -0.00097(7) \\
                 Ir  & Ir$^{4+}$  & 16c & 0          & 0     & 0      & 0.00233(7) & 0.00233(7) & 0.00233(7) &  0.00008(7) &  0.00008(7) &  0.00008(7) \\
                 O1   &  O$^{2-}$ & 48f & 0.33536(3) & 0.125 & 0.125 & 0.00535(9) & 0.00535(9) & 0.00535(9) &  0          &  0          &  0.00164(10)\\
                 O2   &  O$^{2-}$ & 8b  & 0.375 & 0.375 & 0.375 & 0.00340(19) & 0.00340(19) & 0.00340(19) &  0          &  0          &  0 \\

  \hline
\end{tabular}
\caption{Table of refined structural parameters for the $T=290$~K measurement of Y$_2$Ir$_2$O$_7$.  Measurements were made using the HB2A diffractometer.  $a$ is the cubic lattice constant.}
\label{tab:paramtab}
\end{table*}

\begin{figure}[!h]
\centerline{\includegraphics[scale=.5]{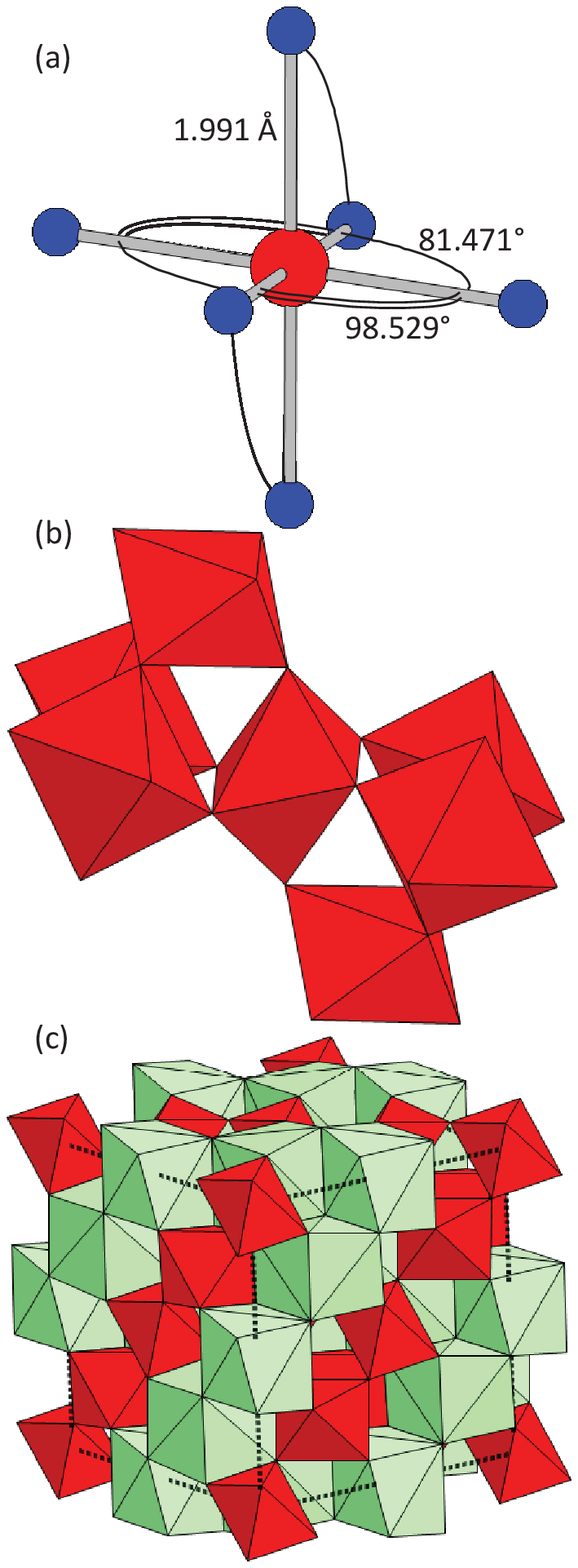}}
\caption{\label{fig:structurefig}
(color online) (a) Local environment of distorted octahedra of the Ir site in Y$_2$Ir$_2$O$_7$.  Large red sphere represents the Ir site and small blue spheres the O sites.  (b)  Corner shared Ir-O octahedra in Y$_2$Ir$_2$O$_7$.  (c)  Unit cell of Y$_2$Ir$_2$O$_7$ showing Ir-O octahedra (red) and interleaving Y-O distorted dodecahedra (light green).  Dashed line is the cubic unit cell.}
\end{figure}

Rietveld structural refinement was performed for the HB2A data using the \textit{FullProf Suite}. \cite{fullprof}  The pyrochlore material Y$_2$Ru$_2$O$_7$ was used as the starting structure, space group cubic $FD\bar{3}M$, for the refinement at $T = 290$ K. \cite{Y2Ru2O7ref}  Variables in the refinements included the lattice constant, the O2 $x$ fractional coordinate, and the anisotropic thermal parameters.  Figure \ref{fig:Ivstwotheta} illustrates the $T = 290$ K measurement and the result of the Rietveld refinement.  The $T = 290$ K cubic lattice constant was refined as 10.10580(7) {\AA} and the refined structural parameters are listed in Table \ref{tab:paramtab}.  The aluminum of the sample holder was included in the refinement as an additional phase.  The green asterisks in Figure \ref{fig:Ivstwotheta} show the positions of an unindexed impurity phase observed in all the measurements. Figure \ref{fig:structurefig} illustrates different portions of the crystal structure.  The local environment of the Ir ion, Figure \ref{fig:structurefig}(a), is a distorted octahedron with all Ir-O distances 1.991 {\AA} and alternating 81.471$^{\circ}$ and 98.529$^{\circ}$ O-Ir-O bonds.  The Ir octahedra are corner sharing in a typical pyrochlore fashion with interleaving Y dodecahedra as shown in Figure \ref{fig:structurefig}.

\begin{figure}[!h]
\centerline{\includegraphics[width=2.9in]{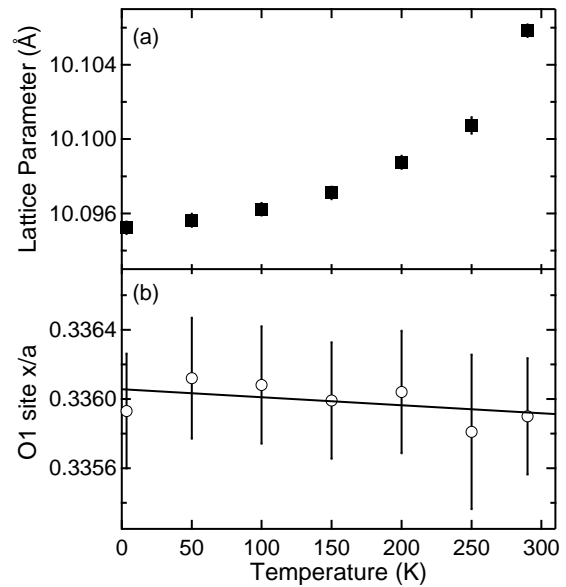}}
\caption{\label{fig:latticeTdepend}
(a) Refined lattice constant of Y$_2$Ir$_2$O$_7$ as a function of temperature. (b) Refined O1 site $x/a$ fractional coordinate as a function of temperature. Results are based upon refinements of HB2A diffraction measurements.  Solid line in (b) is a linear fit as described in the text.}
\end{figure}

The structural refinement was also performed for the other temperatures measured using the HB2A diffractometer.  The anisotropic thermal parameters did not change significantly with temperature.  The lattice constant and the O1 fractional coordinate $x/a$ are shown as a function of temperature in Figure \ref{fig:latticeTdepend}.  There is a non-linear dependent expansion of the lattice above $T \approx 200$ K, potentially signaling anharmonicity in the crystal lattice.  The O1 site fractional coordinate is nearly independent of temperature.  We fit these data to a line with an intercept of 0.33606(6) {\AA} and a small slope ($-5(4) \times 10^{-7}$ {\AA}/K).  No additional measurable reflections were seen as the temperature was reduced.  Likewise no increase of nuclear reflection intensity was observed which could be considered from a $Q = 0$ magnetic ordering.

\begin{figure}[!h]
\centerline{\includegraphics[width=3.5in]{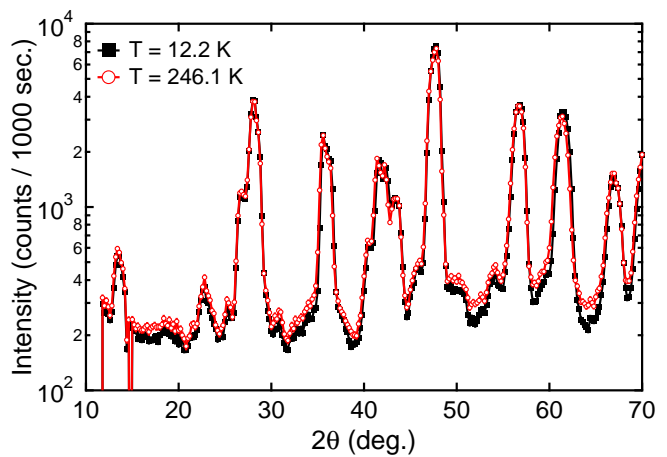}}
\caption{\label{fig:wanddata}
(color online) Temperature dependent diffraction measurements using the WAND diffractometer.  Data are only shown for scattering angles less than 70 degrees.  No clear indication of any short- or long-range magnetic ordering is observed, as described in the text.}
\end{figure}

Figure \ref{fig:wanddata} illustrates the results of two measurements using the WAND diffractometer.  Again, no additional Bragg peaks are observed below $T_{mag}$.  No significant increase in scattering intensity of nuclear Bragg peak positions is observed as a function of decreasing temperature.  We do not observe any clear indication of an antiferromagnetic ($Q \neq 0$) or a $Q = 0$ magnetic long-range ordered phase.  Furthermore, we do not observe any broad increase in scattering intensity at lower temperatures, which may be indicative of glassiness in the magnetic dynamics.  Examining both the WAND and the HB2A diffraction results, we are able to place an upper bound on the ordered magnetic moment resulting from $Q \neq 0$ bragg peaks of $\sim$ 0.2 $\mu_B /$Ir atom and a larger detection limit of $\sim$ 0.5 $\mu_B /$Ir for additional intensity at nuclear bragg peaks resulting from $Q = 0$ ordering.  These values are based upon the statistical uncertainty in the measurements and the magnetic form factor of the iridium ion, including the known iridium cross section and the geometry of the container.

\begin{figure}[!h]
\centerline{\includegraphics[width=3.25in]{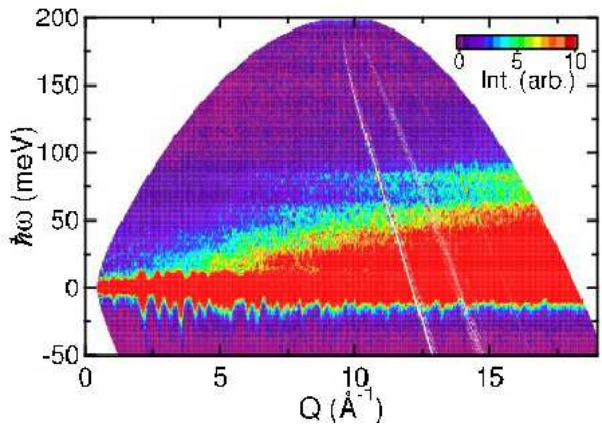}}
\caption{\label{fig:arcsslice}
(color online) Inelastic neutron scattering spectra of Y$_2$Ir$_2$O$_7$ measured with the ARCS instrument with $E_i = 200$ meV at $T = 5$ K.  Data have not been background subtracted with an empty sample can measurement.}
\end{figure}

\begin{figure}[!h]
\centerline{\includegraphics[width=3.5in]{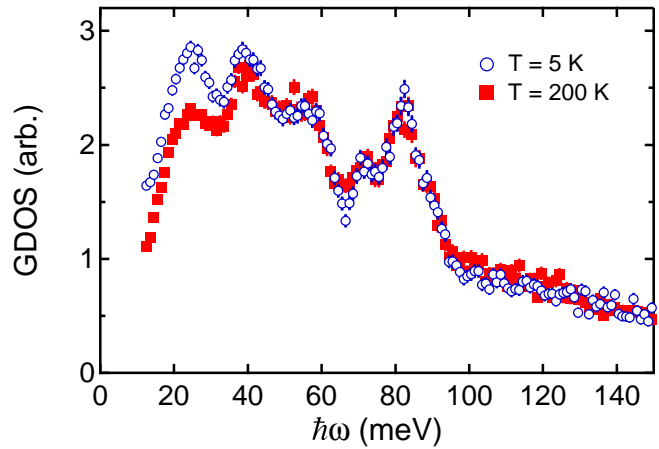}}
\caption{\label{fig:gdos}
(color online) General density of states (GDOS) of Y$_2$Ir$_2$O$_7$ measured with the ARCS instrument with $E_i=200$~meV.  Data have not been background subtracted with an empty sample can measurement.}
\end{figure}

Time-of-flight inelastic neutron scattering measurements were performed with $E_i =$ 3 and 12 meV using the CNCS instrument and $E_i =$ 35, 200, 500 and 2000 meV using the ARCS instrument.  The elastic channel of these measurements was examined for any short or long-range magnetic order as a function of temperature.  No significant signal consistent with magnetism was found in comparing the data above and below the magnetic transition temperature.  The inelastic portion of the spectra were also examined for magnetic excitations or crystal field levels.  No clear magnetic excitations or crystal field levels were observed up to 2000 meV energy transfer. Figure \ref{fig:arcsslice} shows the intensity as a function of energy transfer, $\hbar\omega$ and $Q$ for $E_i = 200$ meV at $T = 5$ K. There is a clear cut off in the optical phonons in the vicinity of 82.5 meV. The general phonon density of states (GDOS) is calculated from these data and shown in Figure \ref{fig:gdos} for $T = 5$ and $200$ K. \cite{daveproject}  The 82.5 meV peak is likely due to oxygen optic modes in the pyrochlore structure.  We find no change in this or the other vibrational frequencies as a function of temperature, indicating that there are no significant changes in the phonon spectrum as the sample is cooled through the magnetic transition.

\section{CONCLUSION}
Y$_2$Ir$_2$O$_7$ undergoes a magnetic transition at $T_{mag}$ $\approx 155$ K. For temperatures below $T_{mag}$, the magnetization is hysteretic, but with only a small remanent magnetization. Neutron diffraction experiments do not show any evidence for a structural phase transition between 290 K and 3.4 K, nor do they show any signature of the magnetic transition. These measurements do not rule out long range magnetic order, but they do put an upper bound for the ordered iridium moment of $\sim$ 0.2 $\mu_B /$Ir (for a $Q \neq 0$ magnetic structure) or $\sim$ 0.5 $\mu_B /$Ir (for $Q = 0$). Our null results are similar to recent measurements of the analogous iridate pyrochlore Nd$_2$Ir$_2$O$_7$ \cite{tomiyasu11}, but without the additional effects associated with the Nd magnetic sublattice.

\section{ACKNOWLEDGEMENTS}
The research at Oak Ridge National Laboratory's High Flux Isotope Reactor and Oak Ridge National Laboratory's Spallation Neutron Source was sponsored by the Scientific User Facilities Division, Office of Basic Energy Sciences, U. S. Department of Energy (DOE).  Work at Stanford was supported by the U.S. DOE, Office of Basic Energy Sciences, under contract DEAC02-76SF00515.

\end{document}